\documentstyle[prd,aps]{revtex}

\input epsf
\epsfverbosetrue

\def\Journal#1#2#3#4{{#1} {\bf #2} (#4) #3}
\def\MPL{Mod. Phys. Lett. A}

\def\NPB{Nucl. Phys. B}

\def\NPSUPPL{Nucl. Phys. Proc. Suppl.}
\def\PLB{{Phys. Lett.} B}

\def\PLBOLD{Phys. Lett.}
\def\PRL{Phys. Rev. Lett.}
\def\RMP{Rev. Mod. Phys.}
\def\PRD{Phys. Rev. D}

\def\PTP{Prog. Theor. Phys.}
\def\JHEP{JHEP}
\def\EPJ{Euro. Phys. J. C}
\def\JETPUSSR{JETP (USSR)}
\def\ZETP{Zh. Eksp. Teor. Piz.}
\def\TNYAS{Trans. New York Acad. Sci.}

\def\mapgeq{\mathbin{\lower.3ex\hbox{$\buildrel>\over{\smash{\scriptstyle\sim}\vphantom{_x}}$}}}
\def\mapleq{\mathbin{\lower.3ex\hbox{$\buildrel<\over{\smash{\scriptstyle\sim}\vphantom{_x}}$}}}
\def\mapgeqeq{\mathbi{\lower.3ex\hbox{$\buildrel>\over{\smash{\scriptstyle\approx}\vphantom{_2}}$}}}
\def\mapleqeq{\mathbin{\lower.3ex\hbox{$\buildrel<\over{\smash{\scriptstyle\approx}\vphantom{_2}}$}}}

 \mathchardef\#="0023
 \mathchardef\$="0024
 \mathchardef\%="0025
 \mathchardef\ddash="705C
 
 \mathchardef\lwavy="336E
 \mathchardef\rwavy="336F
 \mathchardef\biglwavy="331A
 \mathchardef\bigrwavy="331B
 \mathchardef\bigglwavy="3328
 \mathchardef\biggrwavy="3329
 \mathchardef\littlesum="0350

\tighten
\draft
\begin{document} 
\bibliographystyle{prsty}

\title{
Large Solar Neutrino Mixing and Radiative Neutrino Mechanism 
}

\author{
Teruyuki Kitabayashi$^a$
\footnote{E-mail:teruyuki@post.kek.jp}
and Masaki Yasu${\grave {\rm e}}^b$
\footnote{E-mail:yasue@keyaki.cc.u-tokai.ac.jp}
}

\address{\vspace{5mm}$^a$
{\sl Accelerator Engineering Center} \\
{\sl Mitsubishi Electric System \& Service Engineering Co.Ltd.} \\
{\sl 2-8-8 Umezono, Tsukuba, Ibaraki 305-0045, Japan}
}
\address{\vspace{2mm}$^b$
{\sl Department of Natural Science\\School of Marine
Science and Technology, Tokai University}\\
{\sl 3-20-1 Orido, Shimizu, Shizuoka 424-8610, Japan\\and\\}
{\sl Department of Physics, Tokai University} \\
{\sl 1117 KitaKaname, Hiratsuka, Kanagawa 259-1292, Japan}}
\date{TOKAI-HEP/TH-0106, October, 2001}
\maketitle

\begin{abstract}
We find that the presence of a global $L_e-L_\mu-L_\tau$ ($\equiv L^\prime$) symmetry and an $S_2$ permutation symmetry for the $\mu$- and $\tau$-families supplemented by a discrete $Z_4$ symmetry naturally leads to almost maximal atmospheric neutrino mixing and large solar neutrino mixing, which arise, respectively, from type II seesaw mechanism initiated by an $S_2$-symmetric triplet Higgs scalar $s$ with $L^\prime=2$ and from radiative mechanism of the Zee type initiated by two singly charged scalars, an $S_2$-symmetric $h^+$ with $L^\prime=0$ and an $S_2$-antisymmetric $h^{\prime +}$ with $L^\prime=2$.  The almost maximal mixing for atmospheric neutrinos is explained by the appearance of the democratic coupling of $s$ to neutrinos ensured by $S_2$ and $Z_4$ while the large  mixing for solar neutrinos is explained by the similarity of $h^+$- and $h^{\prime +}$-couplings described by $f^h_+\sim f^h_-$ and $\mu_+\sim\mu_-$, where $f^h_+$ ($f^h_-$) and $\mu_+$ ($\mu_-$) stand for $h^+$ ($h^{\prime +}$)-couplings, respectively, to leptons and to Higgs scalars.
\end{abstract}
\vspace{4mm}

Neutrino oscillations have been long recognized to occur if neutrinos have masses \cite{EarlyMassive}.  The experimental confirmation of such neutrino oscillations has been given by the Super-Kamiokande collaboration \cite{Kamiokande} for atmospheric neutrinos and the clear evidence of the solar neutrino oscillations has been released by the SNO collaboration \cite{SNO}.  These observed oscillation phenomena can be explained by the mixings between $\nu_e$ and $\nu_\mu$ with $\Delta m^2_\odot \mapleq 10^{-4}$ eV$^2$ for solar neutrinos and between $\nu_\mu$ and $\nu_\tau$ with $\Delta m^2_{atm} \sim 3 \times 10^{-3}$ eV$^2$ for atmospheric neutrinos \cite{K2K}.  Their masses are implied to be as small as ${\mathcal{O}}(10^{-2})$ eV and the smallness of neutrino masses can be explained by either the seesaw mechanism \cite{Seesaw} or the radiative mechanism \cite{Zee,Babu}.  The mixing specific to atmospheric neutrinos is found to prefer maximal mixing \cite{RecentAtmAnalysis}.  It has also been suggested for solar neutrinos that solutions with large mixing angles are favored while solutions with small mixing angles are disfavored \cite{RecentSK}. Therefore, both neutrino oscillations are characterized by large neutrino mixings.

One of the promising theoretical assumptions to account for the observed mixing pattern is to use the bimaximal mixing scheme \cite{BimaximalMixing,NearlyBiMaximal}.  The radiative mechanism of the Zee-type \cite{Zee} provides the natural explanation on bimaximal neutrino mixing \cite{Useful} when combined with a global $L_e-L_\mu-L_\tau$ ($\equiv L^\prime$) symmetry \cite{EarlierLprime,Lprime} since the Zee model only supplies flavor-off-diagonal mass terms.  However, the recent extensive analyses on solar neutrino oscillation data imply that the maximal solar neutrino mixing is not well compatible with the data, which prefer $\sin^22\theta_{12} \sim 0.8$ for the large mixing angle (LMA) MSW solution \cite{RecentSolarAnalysis}.  If this observed tendency of solar neutrino oscillations with large mixing but not with maximal mixing is really confirmed, the bimaximal structure in the Zee model should be modified \cite{ZeeMaximal}.  

In this report, we discuss a possible modification of the Zee model with the $L^\prime$ symmetry to accommodate the LMA solution without the maximal solar neutrino mixing \cite{Previous}.  The original Zee model requires the presence of a Higgs scalar of $\phi^\prime$, the duplicate of the standard Higgs scalar $\phi$, which initiates radiative neutrino mechanism together with a singly charged scalar of $h^+$, and assumes $\phi^\prime$ to couple to no leptons. One of the modifications is to relax this constraint such that $\phi^\prime$ couples to leptons.  By allowing $\phi^\prime$ to generate lepton masses, the authors of Ref.\cite{ZeeLMA} have found that solar neutrinos can exhibit $\sin^22\theta_{12} \sim 0.8$ but their realization of the large solar neutrino mixing entails various fine-tunings, which seem unnatural.  We, instead, rely upon a certain underlying symmetry to constrain the interactions of $\phi^\prime$ with leptons and utilize an $S_2$ permutation symmetry for the $\mu$- and $\tau$-families, which is responsible for the appearance of the almost maximal atmospheric neutrino mixing \cite{DiscreteSym}.  Under $S_2$, $\phi$ transforms as a symmetric state and $\phi^\prime$ transforms as an antisymmetric state.

In order to realize the large mixing, the natural resolution is to include flavor-diagonal mass terms because the main source of $\sin^22\theta_{12} \approx 1$ in the Zee model comes from the constraint on neutrino masses of $m_{1,2,3}$ given by $m_1+m_2+m_3$=0 specific to flavor-off-diagonal mass terms.  It is known that flavor-diagonal mass terms can be supplied by an $SU(2)_L$-triplet Higgs scalar \cite{Triplet} denoted by
\begin{eqnarray}\label{Eq:s-Higgs}
&s=\left( \begin{array}{cc}
  s^+ &  s^{++}\\
  s^0 &  -s^+
\end{array} \right),
\end{eqnarray}
whose vacuum expectation value (VEV), $\langle 0 \vert s^0 \vert 0 \rangle$, generates neutrino masses via interactions of ${\overline {\psi^c_L}} s \psi_L$, where the subscript $c$ denotes the charge conjugation including the $G$-parity of $SU(2)_L$. The smallness of the neutrino masses can be ascribed to that of $\langle 0 \vert s^0 \vert 0 \rangle$, which is given by $\sim \mu (\langle 0 \vert \phi \vert 0 \rangle/m_s)^2$ produced by the combined effects of $\mu\phi^\dagger s \phi^c$ and $m^2_s$Tr$(s^\dagger s)$, where $\mu$ and $m_s$ are mass parameters. The type II seesaw mechanism \cite{Type2SeeSaw} can ensure tiny neutrino masses by the dynamical requirement of $\vert \langle 0 \vert \phi \vert 0 \rangle\vert \ll m_s$ with $\mu\sim m_s$.

To see which masses of flavor neutrinos give contributions to yield $\sin^22\theta_{12}\neq 1$, we examine a possible neutrino mass texture that can be diagonalized by $U_{MNS}$ with two mixing angles, $\theta_{12}$ and $\theta_{23}$, which, respectively, connect ($\nu_1$, $\nu_2$) with ($\nu_e$, $\nu_\mu$) and ($\nu_2$, $\nu_3$) with ($\nu_\mu$, $\nu_\tau$), where ($\nu_1, \nu_2, \nu_3$)$^T$ (= $\vert \nu_{mass}\rangle$) with ($m_1, m_2, m_3$) and ($\nu_e, \nu_\mu, \nu_\tau$)$^T$ (=$\vert \nu_{weak}\rangle$) are related by $\vert \nu_{weak}\rangle = U_{MNS} \vert \nu_{mass}\rangle$. The resulting mass matrix denoted by $M^\nu$ takes the form of
\begin{eqnarray}\label{Eq:M_NU}
&M^\nu = \left( \begin{array}{ccc}
  a           &  b     & c (= -t_{23}b)\\
  b           &  d     & e\\
  c      &  e     & f(=d+\left( t^{-1}_{23}-t_{23}\right) e)
\end{array} \right),
\end{eqnarray}
where the atmospheric neutrino mixing is specified by $t_{23}$ = $\sin\theta_{23}/\cos\theta_{23}$ \cite{ZeeLMA,DiscreteSym,MassMatrix}.  The masses and the solar neutrino mixing angle of $\theta_{12}$ are calculated to be:
\begin{eqnarray}
&&
m_1 = a-\frac{1}{2}\sqrt{\frac{b^2+c^2}{2}}\left( x+\eta\sqrt{x^2 + 8}\right), \quad m_2 = \left( \eta \rightarrow -\eta ~{\rm in}~m_1\right), 
\nonumber \\
&&
m_3 = d + t^{-2}_{23}\left( d-a +x\sqrt{\frac{b^2+c^2}{2}}\right),
\label{Eq:NuMasses123} \\
&&
\sin^22\theta_{12}=\frac{8}{8+x^2}~{\rm with}~x=\frac{a-d+t_{23}e}{\sqrt{(b^2+c^2)/2}},
\label{Eq:SquareAngle12}
\end{eqnarray}
where $c=-t_{23}b$ and $\vert m_1\vert < \vert m_2\vert$ is always maintained by adjusting the sign of $\eta$ (= $\pm 1$). The result shows that the significant deviation of $\sin^22\theta_{12}$ from unity is only possible if $(a-d+t_{23}e)^2 = {\cal O}(b^2+c^2)$. In our subsequent discussions, we take the ``ideal" solution \cite{Moha} with $t_{23}=\pm 1$ ($\equiv\sigma$) given by
\begin{eqnarray}\label{Eq:SimplestNuMass}
&M^\nu_{ideal}=
\left( \begin{array}{ccc}
  0            &  0            & 0\\
  0            &  d            & \sigma d\\
  0    &  \sigma d     & d
\end{array} \right),
\end{eqnarray}
which provides $m_1=m_2 =0$ and $m_3 = 2d$.  The deviation from this solution that yields $b\neq 0$ and $c\neq 0$ is caused by radiative effects, which also add additional contributions to $d$ and $e$, and $(d-t_{23}e)^2 = {\cal O}(b^2+c^2)$ is ensured by $S_2$. Then, the splitting between $m_1$ and $m_2$ is induced to yield the LMA solution with $\sin^22\theta_{12}\sim 0.8$.

To realize the ``ideal" solution, we introduce a permutation symmetry, $S_2$, for the $\mu$- and $\tau$-families \cite{Permutation} as have been announced, which is compatible with the requirement of $L^\prime$. For $s$ with $L^\prime$ = 2, $s$ only couple to the $\mu$- and $\tau$-families, which provide an $S_2$-symmetric democratic mass texture \cite{Democratic} for $\nu_{\mu,\tau}$ with an additional $Z_4$ discrete symmetry to ensure the ``ideal" structure of Eq.(\ref{Eq:SimplestNuMass}), leading to one massless neutrino ($\nu_2$) and one massive neutrino ($\nu_3$).  These two neutrinos radiatively mixed with $\nu_e$ finally give observed neutrino mixings.  

All interactions are taken to conserve $L^\prime$ and to be invariant under the transformation of $S_2$ as well as $Z_4$. The scalars of $\phi$, $s$ and $h^+$ are assigned to $S_2$-symmetric states.  The other scalar of $\phi^\prime$ is assigned to an $S_2$-antisymmetric state and we introduce an additional copy of $\phi^{\prime}$ and $h^+$ as $S_2$-antisymmetric states denoted by $\phi^{\prime\prime}$ and $h^{\prime +}$. The inclusion of $\phi^{\prime\prime}$ and $h^{\prime +}$, respectively, allows us to meet the mass hierarchy of $m_\mu\ll m_\tau$ and the large solar neutrino mixing satisfying $(d-t_{23}e)^2 = {\cal O}(b^2+c^2)$.  To distinguish these copies from the original fields, it is sufficient to use a discrete symmetry of $Z_4$.  The quantum numbers of the participating fields are tabulated in TABLE \ref{Tab:QuantumNumber1}, where $\psi_{\pm L}$ = $(\psi^\tau_L \pm \psi^\mu_L)/\sqrt{2}$ and $\ell_{\pm R}$ = $(\tau_R \pm \mu_R)/\sqrt{2}$. The assignment of the $Z_4$-charges of $\psi_{\pm L}$ and $s$ forbids the coexistence of $\psi_{+ L}s\psi_{+ L}$ and $\psi_{- L}s\psi_{- L}$, which disturbs the democratic structure of the ``ideal" solution.  The present assignment corresponds to the $\sigma = 1$ solution of Eq.(\ref{Eq:SimplestNuMass}).  Since charged leptons simultaneously couple to the Higgs scalars of $\phi$, $\phi^\prime$ and $\phi^{\prime\prime}$, flavor-changing interactions are induced by the exchanges of these Higgs scalars, which will be shown to give well-suppressed contributions at the phenomenologically consistent level. It is obvious that quarks that are $S_2$-symmetric can have couplings to $\phi$ but not to $\phi^\prime$ and $\phi^{\prime\prime}$; therefore, quarks do not have this type of dangerous flavor-changing interactions.

\begin{figure}
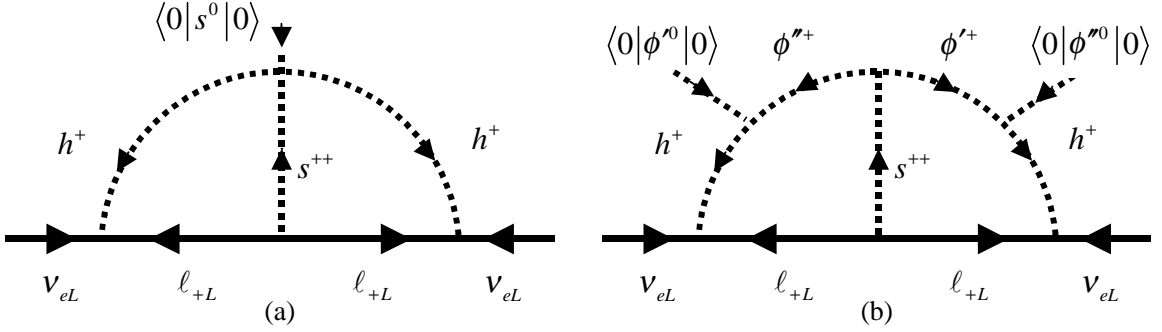

\caption{(a) Divergent two-loop diagram for Majorana mass terms of $\nu_{eL}\nu_{eL}$. (b) The same as (a) but for the finite diagram.}
\label{Fig:TwoLoop1}
\vspace{-7mm}
\end{figure}

The Yukawa interactions for leptons are, now, given by
\begin{eqnarray}
-{\cal L}_Y   &=& 
f_\phi {\overline {\psi^e_L}}\phi e_R
+
{\overline {\psi_{+ L}}}\left( f_+\phi \ell_{+ R} + f_-\phi^\prime \ell_{- R}\right)
+
{\overline {\psi_{- L}}}\left( g_+\phi \ell_{- R} + g_-\phi^{\prime\prime} \ell_{+ R}\right)\nonumber \\
&&
+
f^h_+{\overline {\left( \psi_L^e \right)^c}}\psi_{+ L}h^+
+
f^h_-{\overline {\left( \psi_{+ L} \right)^c}}\psi_{- L}h^{\prime +}
+
f^s_+{\overline {\left( \psi_{+ L} \right)^c}}s\psi_{+ L}
+ {\rm (h.c.)},
\label{Eq:OurYukawa}
\end{eqnarray}
where $f$'s stand for coupling constants.  Higgs interactions are described by usual Hermitian terms composed of $\varphi\varphi^\dagger$ ($\varphi$ = $\phi$, $\phi^\prime$, $\phi^{\prime\prime}$, $h^+$, $h^{\prime +}$, $s$) and by non-Hermitian terms in
\begin{eqnarray}
V_0
& = &
\left( 
\lambda_1\phi^{\prime\prime\dagger} s\phi
+\lambda_2\phi^\dagger s\phi^\prime
\right) h^{\prime +\dagger}
+{\rm (h.c.)}, 
\label{Eq:Conserved}
\end{eqnarray}
where $\lambda_{1,2}$ are Higgs couplings, which conserves $L$ and $L^\prime$.  The soft breaking terms of $L$ and $L^\prime$ can be chosen to be:
\begin{eqnarray}
&&
V_1
=
\mu_+\phi^{\prime c \dagger}\phi^{\prime\prime}h^{+ \dagger}+{\rm (h.c.)}, 
\quad
V_2
=
\mu_-\phi^{c \dagger}\phi^{\prime\prime}h^{\prime + \dagger}
+{\rm (h.c.)}, 
\nonumber \\
&&
V_3
=
\mu\phi^\dagger s \phi^c
+
\mu^\prime\phi^{\prime\dagger} s \phi^{\prime\prime c}
+{\rm (h.c.)}, 
\label{Eq:Broken1}
\end{eqnarray}
where $\mu$'s represent mass scales and $V_{1,2}$ and $V_3$ are, respectively, used to activate the radiative mechanism and the type II seesaw mechanism.  Although $L$ and $L^\prime$ are spontaneously broken by $\langle 0 \vert s \vert 0\rangle$, $L$+$L^\prime$ ($\propto L_e$) is still conserved.  In terms of the $L_e$-conservation, $V_2$ and $V_3$ are classified as $L_e$-conserving interactions and its explicit breaking is provided by $V_1$. $L_e$-breaking interactions such as those causing $\mu,\tau\rightarrow eee$ and $\rightarrow e\gamma$ necessarily involve $V_1$.  All other interactions are forbidden by the conservation of $L_e$ and $Z_4$.  Especially, $(h^+h^+)^\dagger \det s$ could give a divergent term of $\nu_e\nu_e$ at the two loop level as depicted in FIG.\ref{Fig:TwoLoop1} (a), which then would require a tree level mass term of the $\nu_e\nu_e$-term as a counter term. The appearance of this counter term is not consistent with the absence of the tree-level $\nu_e\nu_e$-term in Eq.(\ref{Eq:SimplestNuMass}).  Since $L^\prime$ is explicitly broken, $\nu_e\nu_e$ is induced by interactions shown in FIG.\ref{Fig:TwoLoop1} (b) with two $V_1$ insertions.  Fortunately, this diagram leads to the finite convergent term.

Charged lepton masses are generated via the Higgs couplings to leptons, which are specified by the following matrix of $M^\ell_0 (\phi, \phi^\prime, \phi^{\prime\prime})$:
\begin{eqnarray}\label{Eq:OUR_M_ELL}
&&M^\ell_0\left( \phi,\phi^\prime, \phi^{\prime\prime}\right)
=
\left( \begin{array}{ccc}
f_\phi \phi 
&
0
&
0
\\
0
&
\frac{1}{2}\left[
\left(f_++g_+\right)\phi
-
\left(f_-\phi^\prime + g_-\phi^{\prime\prime}\right)
\right]
&
\frac{1}{2}\left[
\left(f_+-g_+\right)\phi
+
\left(f_-\phi^\prime - g_-\phi^{\prime\prime}\right)
\right]
\\
0
&
\frac{1}{2}\left[
\left(f_+-g_+\right)\phi
-
\left(f_-\phi^\prime - g_-\phi^{\prime\prime}\right)
\right]
&
\frac{1}{2}\left[
\left(f_++g_+\right)\phi
+
\left(f_-\phi^\prime + g_-\phi^{\prime\prime}\right)
\right]
\end{array} \right).
\end{eqnarray}
The masses for leptons denoted by $M^\ell_0$ are, thus, described by
\begin{eqnarray}\label{Eq:OUR_M_ELL1}
&
M^\ell_0
= \langle 0 \vert M^\ell_0 (\phi, \phi^\prime, \phi^{\prime\prime}) \vert 0 \rangle
=
\left( \begin{array}{ccc}
  m_e   & 0        & 0\\
  0     & m^\ell_{\mu\mu}  & m^\ell_{\mu\tau}\\
  0     & m^\ell_{\tau\mu}& m^\ell_{\tau\tau}
\end{array} \right)
\end{eqnarray}
with
\begin{eqnarray}\label{Eq:OUR_M_ELL2}
&&
m_e = f_\phi v,
\quad
m^\ell_{\mu\mu} = \frac{1}{2}\left[
\left(f_++g_+\right)v
-
\left(f_-v^\prime + g_-v^{\prime\prime}\right)
\right],
\quad
m^\ell_{\tau\tau} = \frac{1}{2}\left[
\left(f_++g_+\right)v
+
\left(f_-v^\prime + g_-v^{\prime\prime}\right)
\right],
\nonumber \\
&&
m^\ell_{\mu\tau} = \frac{1}{2}\left[
\left(f_+-g_+\right)v
+
\left(f_-v^\prime - g_-v^{\prime\prime}\right)
\right],
\quad
m^\ell_{\tau\mu} = \frac{1}{2}\left[
\left(f_+-g_+\right)v
-
\left(f_-v^\prime - g_-v^{\prime\prime}\right)
\right],
\end{eqnarray}
where $v$ = $\langle 0 \vert \phi^0 \vert 0 \rangle$, $v^\prime$ = $\langle 0 \vert \phi^{\prime 0} \vert 0 \rangle$ and $v^{\prime\prime}$ = $\langle 0 \vert \phi^{\prime\prime 0} \vert 0 \rangle$.  To be consistent with the pattern of the observed hierarchy of  $m_e\ll m_\mu\ll m_\tau$, we simply adopt the parameterization based on the ``hierarchical" one \cite{Hierarchical}.

It is straight forward to reach $U_\ell$ ($V_\ell$) that links the original states of $\vert \ell^0_{L(R)} \rangle$ to the states with the diagonal masses of $\vert \ell_{L(R)} \rangle$: $\vert \ell^0_L \rangle = U_\ell \vert \ell_L \rangle$ and $\vert \ell^0_R \rangle = V_\ell \vert \ell_R \rangle$.  The original mass matrix $M^\ell_0$ is transformed into $M^\ell$ according to $M^\ell=U^\dagger_\ell M^\ell_0 V_\ell$=diag.($m_e$, $m_\mu$, $m_\tau$):
\begin{eqnarray}\label{Eq:Unitary}
&
U_\ell=\left( \begin{array}{ccc}
1  & 0 & 0\\
0  & c_\alpha & s_\alpha\\
0 & -s_\alpha & c_\alpha
\end{array} \right),
\quad
V_\ell=\left( \begin{array}{ccc}
1  & 0 & 0\\
0  & c_\beta & s_\beta\\
0 & -s_\beta & c_\beta
\end{array} \right),
\end{eqnarray}
where $c_\alpha = \cos\alpha$ etc. defined by 
\begin{eqnarray}
&&
c_\alpha = \sqrt{
\frac
{m^{\ell 2}_{\tau\tau}+m^{\ell 2}_{\tau\mu}-m^2_\mu}
{m^2_\tau-m^2_\mu}
},
\quad
s_\alpha = \sqrt{
\frac
{m^2_\tau-m^{\ell 2}_{\tau\tau}-m^{\ell 2}_{\tau\mu}}
{m^2_\tau-m^2_\mu}
},
\label{Eq:Alpha}\\
&&
c_\beta = \sqrt{
\frac
{m^{\ell 2}_{\tau\tau}+m^{\ell 2}_{\mu\tau}-m^2_\mu}
{m^2_\tau-m^2_\mu}
},
\quad
s_\beta = \sqrt{
\frac
{m^2_\tau-m^{\ell 2}_{\tau\tau}-m^{\ell 2}_{\mu\tau}}
{m^2_\tau-m^2_\mu}
}.
\label{Eq:Beta}
\end{eqnarray}
The $\mu$ and $\tau$ masses are calculated to be $m^2_\mu = \lambda_-$ and $m^2_\tau = \lambda_+$ with $\lambda_\pm$ given by
\begin{eqnarray}
\lambda_\pm &=& \frac{1}{2}
\left(
m^{\ell 2}_{\tau\tau}
+
m^{\ell 2}_{\mu\mu}
+
m^{\ell 2}_{\tau\mu}
+
m^{\ell 2}_{\mu\tau}
\pm
M^2
\right),
\label{Eq:MuTauMasses}
\end{eqnarray}
where
\begin{eqnarray}
M^4 &=& 
\left(
m^{\ell 2}_{\tau\tau}
-m^{\ell 2}_{\mu\mu}
\right)^2
+
\left(
m^{\ell 2}_{\tau\mu}
-m^{\ell 2}_{\mu\tau}
\right)^2
+
2\left(
m^\ell_{\tau\tau}m^\ell_{\mu\tau}
+
m^\ell_{\mu\mu}m^\ell_{\tau\mu}
\right)^2
+
2\left(
m^\ell_{\tau\tau}m^\ell_{\tau\mu}
+
m^\ell_{\mu\mu}m^\ell_{\mu\tau}
\right)^2.
\label{Eq:MuTauMasses2}
\end{eqnarray}
The hierarchical mass pattern of $m_\mu\ll m_\tau$ can be realized by the hierarchical conditions of $\vert s_\alpha\vert,\vert s_\beta\vert \ll 1$.  It is convenient for our later discussions to relate $m^\ell_{ij}$ with $m_{\mu,\tau}$, which are described by
\begin{eqnarray}
&&
m^\ell_{\mu\mu}
=
S^2m_\tau+C^2m_\mu,
\quad
m^\ell_{\tau\tau}
=
C^2m_\tau+S^2m_\mu,
\nonumber \\
&&
m^\ell_{\mu\tau}
=
\frac{1}{c^2_\beta -s^2_\alpha}
\left[
\left(
c_\alpha s_\alpha C^2-c_\beta s_\beta S^2
\right)
m_\tau
-
\left(
c_\beta s_\beta C^2-c_\alpha s_\alpha S^2
\right)
m_\mu
\right],
\nonumber \\
&&
m^\ell_{\tau\mu}
=
\frac{1}{c^2_\beta -s^2_\alpha}
\left[
\left(
c_\beta s_\beta C^2-c_\alpha s_\alpha S^2
\right)
m_\tau
-
\left(
c_\alpha s_\alpha C^2-c_\beta s_\beta S^2
\right)
m_\mu
\right],
\label{Eq:Elements}
\end{eqnarray}
where
\begin{eqnarray}
&&
C^2 = \frac{c^2_\alpha+c^2_\beta}{2},
\quad
S^2 = \frac{s^2_\alpha+s^2_\beta}{2}.
\label{Eq:CombinedSincos}
\end{eqnarray}
By combining Eqs.(\ref{Eq:OUR_M_ELL2}) and (\ref{Eq:Elements}), we find that
\begin{eqnarray}
&&
f_+v \sim g_+v \sim \left( m_\tau+m_\mu\right)/2,
\quad
f_-v^\prime \sim g_-v^{\prime\prime} \sim \left( m_\tau-m_\mu\right)/2,
\label{Eq:Couplings_Constraint}
\end{eqnarray}
should be satisfied for $\vert s_\alpha\vert,\vert s_\beta\vert \ll 1$.

Even after the rotation that gives the diagonal mass matrix of $U^\dagger_\ell\langle 0 \vert M^\ell_0 (\phi, \phi^\prime, \phi^{\prime\prime}) \vert 0 \rangle V_\ell$, our Yukawa interactions corresponding to $U^\dagger_\ell M^\ell_0 (\phi, \phi^\prime, \phi^{\prime\prime}) V_\ell$ (=$M^\ell\left( \phi,\phi^\prime, \phi^{\prime\prime}\right)$) still contain flavor-off-diagonal couplings.  In fact, Eq.(\ref{Eq:OUR_M_ELL1}) is transformed into $M^\ell\left( \phi,\phi^\prime, \phi^{\prime\prime}\right)$, whose elements denoted by $M_{ij}$ are calculated to be:
\begin{eqnarray}\label{Eq:OUR_YUKAWA}
&&
M_{11} = \alpha_\phi m_e,
\quad
M_{12} = M_{21} = M_{13} = M_{31} = 0,
\nonumber \\
&&
M_{22}
=
\frac{m_\tau+m_\mu}{2}\alpha_\phi
-
\frac{m_\tau-m_\mu}{4}\left(\alpha_{\phi^\prime}+\alpha_{\phi^{\prime\prime}}\right)
-
\left( s_\alpha-s_\beta\right)\frac{m_\tau}{2}\left( \alpha_{\phi^\prime}-\alpha_{\phi^{\prime\prime}}\right),
\nonumber \\
&&
M_{23}
=
-
\frac{m_\tau-m_\mu}{4}\left(\alpha_{\phi^\prime}-\alpha_{\phi^{\prime\prime}}\right)
+
\frac{1}{2}\left( s_\alpha m_\tau-s_\beta m_\mu\right)\left( 2\alpha_\phi-\alpha_{\phi^\prime}-\alpha_{\phi^{\prime\prime}}\right),
\nonumber \\
&&
M_{32}
=
\frac{m_\tau-m_\mu}{4}\left(\alpha_{\phi^\prime}-\alpha_{\phi^{\prime\prime}}\right)
+
\frac{1}{2}\left( s_\beta m_\tau-s_\alpha m_\mu\right)\left( 2\alpha_\phi-\alpha_{\phi^\prime}-\alpha_{\phi^{\prime\prime}}\right)
\nonumber \\
&&
M_{33}
=
\frac{m_\tau+m_\mu}{2}\alpha_\phi
+
\frac{m_\tau-m_\mu}{4}\left(\alpha_{\phi^\prime}+\alpha_{\phi^{\prime\prime}}\right)
+
\left( s_\alpha-s_\beta\right)\frac{m_\tau}{2}\left( \alpha_{\phi^\prime}-\alpha_{\phi^{\prime\prime}}\right),
\end{eqnarray}
up to ${\mathcal{O}}(s_{\alpha,\beta})$, where $\alpha_\phi=\phi/v$, $\alpha_{\phi^\prime}=\phi^\prime/v^\prime$ and $\alpha_{\phi^{\prime\prime}}=\phi^{\prime\prime}/v^{\prime\prime}$.  One can readily find that the identification of $\alpha_{\phi^\prime}$ and $\alpha_{\phi^{\prime\prime}}$ with $\alpha_\phi$ corresponding to the case of the standard model gives diagonal interactions, leading to the diagonal lepton masses. The flavor-changing interactions involving $\tau$ and $\mu$ such as $\tau\rightarrow\mu\gamma$ and $\tau\rightarrow\mu\mu\mu$ are roughly controlled by the coupling of $m_i/m_H$ (=$\xi_i$), where $i=e,\tau$ and $m_H$ is a mediating Higgs boson mass.  We find constraints on $\xi_{e,\tau}$ to suppress these interactions to the phenomenologically consistent level, which are given by examining the following typical processes:
\begin{enumerate}
\item for $\tau ^ -   \to\mu^ -  e^ -  e^ + $ mediated by $\phi$, $\vert s_{\alpha,\beta}\xi_\tau\xi_e/m^2_H\vert$ $\mapleq$ $2.1 \times 10^{ - 7}$ GeV$^{ - 2}$ from $B(\tau ^ -   \to \mu^ -  e^ -  e^ + )$ $<$ $1.7 \times 10^{-6}$,
\item for $\tau ^ -  \to\mu^ - \mu^ + \mu^ -$ mediated by $\phi^\prime$ and $\phi^{\prime\prime}$, $\vert \xi_\tau/m_H\vert^2$ $\mapleq$ $2.2 \times 10^{ - 7}$ GeV$^{ - 2}$ from $B(\tau ^ -   \to \mu^ -  \mu^ -  \mu^ + )$ $<$ $1.9 \times 10^{-6}$,
\item for $\tau ^ -  \to\mu^ -    \gamma$ mediated by $\phi^\prime$ and $\phi^{\prime\prime}$, $\vert \xi_\tau/m_H\vert^2$ $\mapleq$ $4.2 \times 10^{ - 6}$ GeV$^{ - 2}$ from $B(\tau ^ -   \to \mu^ -   \gamma )$ $<$ $1.1 \times 10^{-6}$,
\end{enumerate}
where the data are taken form Ref.\cite{Data}. Since $m_H \mapgeq v_{weak}$ is anticipated, where $v_{weak}$ = $( 2{\sqrt 2}G_F)^{-1/2}$=174 GeV for the weak boson masses, $\xi_\tau \sim m_\tau/v_{weak}$ and $\xi_e \sim m_e/v_{weak}$ with $\vert s_{\alpha,\beta}\vert \ll 1$ readily satisfy these constraints.  As stated previously, there are no such Higgs interactions for quarks that only couple to $\phi$. The similar flavor-changing interactions caused by $h^+$ and $h^{\prime +}$ \cite{h_phenom2} are sufficiently suppressed because of the smallness of their-couplings to leptons to be estimated in Eq.(\ref{Eq:CouplingValues}).

The radiative neutrino masses, $\delta m^{rad}_{ij}$, are generated by interactions corresponding to FIG.\ref{Fig:OneLoop1}. Let us denote by $M^{vertex}_0$ the amplitude involving contributions from the vertices connected by the mediating Higgs scalar, either one of $\phi$, $\phi^\prime$ and $\phi^{\prime\prime}$, and kinematical factors due to one-loop contributions denoted by $P$, $P^\prime$ and $P^{\prime\prime}$:
\begin{eqnarray}\label{Eq:Vertex}
&
M^{vertex}_0=\left( \begin{array}{ccc}
P U
&
0
&
0
\\
0
&
P V
-P^\prime V^\prime
-P^{\prime\prime} V^{\prime\prime}
&
P W
+P^\prime V^\prime
-P^{\prime\prime} V^{\prime\prime}
\\
0
&
P W
-P^\prime V^\prime
+P^{\prime\prime} V^{\prime\prime}
&
P V
+P^\prime V^\prime
+P^{\prime\prime} V^{\prime\prime}
\end{array} \right)
\end{eqnarray}
with
\begin{eqnarray}
&&
U = f_\phi\mu_- v^{\prime\prime}{\hat h}^\prime,
\quad
V = \left( f_+ + g_+\right)\mu_- v^{\prime\prime}{\hat h}^\prime,
\quad
W = \left( f_+ - g_+\right)\mu_- v^{\prime\prime}{\hat h}^\prime,
\nonumber \\
&&
V^\prime = f_- \mu_+ v^{\prime\prime}{\hat h},
\quad
V^{\prime\prime} = g_- \left( \mu_+v^\prime{\hat h}+ \mu_-v{\hat h}^\prime\right),
\label{Eq:VertexFields}
\end{eqnarray}
where ${\hat h}$ and ${\hat h}^\prime$ project out the contributions of $h^+$ and $h^{\prime +}$ with ${\hat h}{\hat h}$ = ${\hat h}^\prime{\hat h}^\prime$ = 1 and ${\hat h}{\hat h}^\prime$ = 0.  The $U$-term arises from the interaction of $\mu_-\phi^{c \dagger}\phi^{\prime\prime}h^{\prime + \dagger}$ giving $\mu_-{\hat h}^\prime$ and $\langle 0 \vert \phi^{\prime\prime 0}\vert 0 \rangle$ (=$v^{\prime\prime})$ and of $f_\phi {\overline {\psi^e_L}}\phi e_R
$ giving $f_\phi$ with the mediating $\phi^+$ and $h^{\prime +}$ involved in $P$ and similarly for other terms.  The one-loop factors of $P$'s are defined by
\begin{eqnarray}
&&
P = \frac{1}{16\pi^2}
    \frac{\ln m^2_h-\ln m^2_\phi}{m^2_h-m^2_\phi},
\label{Eq:KineticFactor}
\end{eqnarray}
where $m$'s are masses of the relevant scalars and $m_h$ = $m_{h^+}$ ($m_{h^{\prime +}}$) if $P$'s accompany ${\hat h}$ (${\hat h}^\prime$) in Eq.(\ref{Eq:Vertex}) and similarly for $P^\prime$ with $m^2_\phi\rightarrow m^2_{\phi^\prime}$ and $P^{\prime\prime}$ with $m^2_\phi\rightarrow m^2_{\phi^{\prime\prime}}$,   

By considering the rotation effects due to $U_\ell$ that transforms the original states of $\vert \nu_0 \rangle$ into $\vert \nu_{weak} \rangle$: $\vert \nu_{weak} \rangle$ = $U^\dagger_\ell\vert \nu_0 \rangle$, we find that $\delta m^{rad}_{ij}$ can be parameterized by
\begin{eqnarray}
\delta m^{rad}_{ij} 
&=& 
\left( U^T_\ell {\bf f}  M^\ell_0 M^{vertex\dagger}_0 U_\ell\right)_{ij}
=
\left( {\bf f}^\prime  M^\ell M^{vertex\dagger}\right)_{ij},
\label{Eq:RadMassNu}
\end{eqnarray}
for $\vert\nu_{weak}\rangle$, where ${\bf f}_{ij}$ = $f_{[ij]}$ with $f_{[e\mu]}$ = $f_{[e\tau]}$ = $f^h_+{\hat h}/\sqrt{2}$ and $f_{[\mu\tau]}$ = $f^h_-{\hat h}^\prime$, ${\bf f}^\prime = U^T_\ell{\bf f}U_\ell$ and $M^{vertex} = U^\dagger_\ell M^{vertex}_0 V_\ell$.  The radiative neutrino masses given by $\delta m^\nu_{ii} = 2\delta m^{rad}_{ii}$ and $\delta m^\nu_{ij}$ = $\delta m^{rad}_{ij} + \delta m^{rad}_{ji}$ ($i\neq j$) are calculated to be:
\begin{eqnarray}
&&
\delta m^\nu_{ee} = 0,
\quad
\delta m^\nu_{\mu\mu} = \frac{1}{2}f^h_- r^{\prime\prime -1} P^{\prime\prime}\mu_- m^2_\tau,
\quad
\delta m^\nu_{\tau\tau} = \frac{1}{2}f^h_- r^{\prime\prime -1} P^{\prime\prime}\mu_- m_\mu m_\tau,
\nonumber \\
&&
\delta m^\nu_{e\mu} = \frac{1}{4\sqrt{2}}f^h_+\left( r^{-1}P^\prime-rP^{\prime\prime}\right)\mu_+m^2_\tau,
\quad
\delta m^\nu_{e\tau} = -\frac{1}{4\sqrt{2}}f^h_+\left( r^{-1}P^\prime+rP^{\prime\prime}\right)\mu_+m^2_\tau,
\nonumber \\
&&
\delta m^\nu_{\mu\tau} = -\frac{1}{2}f^h_-\left( r^{\prime\prime}P+\frac{1}{2}r^{\prime\prime -1}P^{\prime\prime}\right)\mu_-m^2_\tau,
\label{Eq:RadMassNuEntries}
\end{eqnarray}
where $r$ = $v^\prime/v^{\prime\prime}$ and $r^{\prime\prime}$ = $v^{\prime\prime}/v$ and we have neglected the non-leading contributions of ${\mathcal{O}}$($s_{\alpha,\beta}$) and ${\mathcal{O}}$($m_{\mu,e}/m_\tau$).  The tree level masses, $m^\nu_{ij}$ ($i,j$ = $\mu,\tau$), are given by the type II seesaw mechanism to be: 
\begin{eqnarray}\label{Eq:TreeMassNuEntries}
&&
m^\nu_{ij} =A_{ij}f^s_+ v_s \approx A_{ij}f^s_+\frac{ \mu v^2+\mu^\prime v^\prime v^{\prime\prime}}{2m^2_s},
\label{Eq:TreeLevelMass}
\end{eqnarray}
for $m_s \sim \mu \sim \mu^\prime \gg v,v^\prime, v^{\prime\prime}$, where $v_s$ = $\langle 0 \vert s^0 \vert 0 \rangle$, $A_{\mu\mu}$ = $(c_\alpha-s_\alpha)^2$ ($\sim$ $1-2s_\alpha$), $A_{\tau\tau}$ = $(c_\alpha+s_\alpha)^2$ ($\sim$ $1+2s_\alpha$) and $A_{\mu\tau}$ = $A_{\tau\mu}$ = $c^2_\alpha-s^2_\alpha$ ($\sim$ 1). Our mass matrix of Eq.(\ref{Eq:M_NU}) has the following mass parameters:
\begin{eqnarray}\label{Eq:MassMatrixElement}
&&
a = 0,
\quad
b = \delta m^\nu_{e\mu},
\quad
c = \delta m^\nu_{e\tau},
\nonumber \\
&&
d = (c_\alpha-s_\alpha)^2m^\nu + \delta m^\nu_{\mu\mu},
\quad
e = (c^2_\alpha-s^2_\alpha)m^\nu + \delta m^\nu_{\mu\tau},
\quad
f = (c_\alpha+s_\alpha)^2m^\nu + \delta m^\nu_{\tau\tau}.
\end{eqnarray}
where we have used the exact expressions for $d$, $e$ and $f$ as far as the tree-level contributions are concerned.  The possible contribution to the mass parameter of $a$ from the two-loop convergent diagram of FIG.\ref{Fig:TwoLoop1} (b) is well suppressed by $m_s$ arising from the propagator of $s$ and does not jeopardize $a$=0.

We are now in a position to estimating various neutrino oscillation parameters. In the course of calculations, we assume for the simplicity that $m^2_{h^+}$ = $m^2_{h^{\prime +}}\gg m^2_\phi= m^2_{\phi^\prime}= m^2_{\phi^{\prime\prime}}$, leading to $P = P^\prime = P^{\prime\prime}$.  The mixing angle $t_{23}$ for the atmospheric neutrino oscillations is computed to be:
\begin{eqnarray}\label{Eq:Requirement1}
&&
t_{23} = \frac{1+r^2}{1-r^2},
\end{eqnarray}
from $t_{23} = -c/b (=-\delta m^\nu_{e\tau}/\delta m^\nu_{e\mu})$.  In the limit of $\delta m^\nu_{\mu\tau}=0$, $t_{23}$ is also given by $t_{23}=(1-t_\alpha )/(1+t_\alpha )$ from $f$ = $d$ + $(t^{-1}_{23}-t_{23})e$, which ensures the almost maximal atmospheric neutrino mixing characterized by $t_{23}\approx 1$ because $\vert s_\alpha\vert\ll 1$ for the hierarchical $\mu$ and $\tau$ mass texture.  To be consistent, we require that $r^2=-t_\alpha$, thereby, $v^{\prime\prime} \gg v^\prime$ for $r^2\ll 1$ leading to $g^2_-/f^2_- \ll 1$ from Eq.(\ref{Eq:Couplings_Constraint}).  By including $\delta m^\nu_{\mu\tau}$, we find that $r^2=-t_\alpha$ is modified into
\begin{eqnarray}\label{Eq:Requirement2}
&&
r^2 = -s_\alpha-\frac{1}{2\left( 1+2r^{\prime\prime 2}\right)}\frac{\delta m^\nu_{\mu\tau}}{m^\nu},
\end{eqnarray}
up to ${\mathcal{O}}$($s_\alpha$), where we have replaced $f^h_-P\mu_-m^2_\tau$ by $\delta m^\nu_{\mu\tau}$ defined in Eq.(\ref{Eq:RadMassNuEntries}).  The mixing angle $\sin^22\theta_{12}$ for the solar neutrino oscillations is given by $\sin^22\theta_{12}$ = $8/(8+x^2)$ of Eq.(\ref{Eq:SquareAngle12}) with $x$ calculated to be:
\begin{eqnarray}\label{Eq:Requirement3}
&&
x 
=
\frac{2m^\nu}{\sqrt{(b^2+c^2)/2}}\frac{ s_\alpha+r^2}{1-r^2}-2\sqrt{2}r\frac{ r^{\prime\prime}+r^{\prime\prime -1}}{(1-r^2)^2}\frac{f^h_-\mu_-}{f^h_+\mu_+} 
=
\frac{1+4r^{\prime\prime 2}-r^2}{(1+2r^{\prime\prime 2})}
\frac{\delta m^\nu_{\mu\tau}}{\delta m^\nu_{e\mu}},
\end{eqnarray}
up to $r^2$ and $\vert s_\alpha\vert$, where we have used the relation of
\begin{eqnarray}\label{Eq:CouplingRatio}
&&
\frac{f^h_-\mu_-}{f^h_+\mu_+} 
=
-
\frac{1}{\sqrt{2}r}\frac{1-r^2}{2r^{\prime\prime }+r^{\prime\prime -1}}
\frac{\delta m^\nu_{\mu\tau}}{\delta m^\nu_{e\mu}}
\end{eqnarray}
supplied by Eq.(\ref{Eq:RadMassNuEntries}).  The tree-level contributions to $x$ involving $s_\alpha$ vanish in Eq.(\ref{Eq:Requirement3}) because of the use of Eq.(\ref{Eq:Requirement2}).  This cancellation is realized by the ``ideal" structure of the tree-level mass terms thanks to the presence of $S_2$ and $Z_4$ and can be traced back to the fact that the tree-level contributions alone give $x$=0 since the relations of $t_{23}$ = $(1 - t_\alpha)/(1 + t_\alpha)$ and $x\propto -(c_\alpha - s_\alpha)^2+t_{23}(c^2_\alpha - s^2_\alpha)$ yields $x=0$.  The masses of neutrinos that of course depend on $x$ satisfy the ``normal" mass hierarchy of $\vert m_1\vert < \vert m_2\vert \ll m_3$ determined by Eq.(\ref{Eq:NuMasses123}) to be:
\begin{eqnarray}\label{Eq:EstimatedNuMasses}
&&
m_1 \approx -\eta\frac{1}{2}\delta m^\nu_{rad}\left( \sqrt{x^2 + 8}-\vert x\vert\right), 
\quad
m_2 \approx \eta\frac{1}{2}\delta m^\nu_{rad}\left( \sqrt{x^2 + 8}+\vert x\vert\right), 
\nonumber \\
&&
m_3 \approx 2(1-2s_\alpha)m^\nu + \delta m^\nu_{\mu\mu}+x\delta m^\nu_{rad}
\end{eqnarray}
with $\delta m^\nu_{rad} = \sqrt{\delta m^{\nu 2}_{e\mu}+\delta m^{\nu 2}_{e\tau}}$, where $\eta$ is chosen such that $\eta x = -\vert x \vert$.  Then, $\Delta m^2_{atm,\odot}$ are calculated to be:
\begin{eqnarray}\label{Eq:Sin_12}
&&
\Delta m^2_{atm} = m^2_3-m^2_2 \approx 4m^{\nu 2}
+
4m^\nu
\left(
\delta m^\nu_{\mu\mu}+x\vert\delta m^\nu_{rad}\vert-4s_\alpha m^\nu
\right),
\nonumber \\
&&
\Delta m^2_\odot = m^2_2-m^2_1 \approx \vert x\vert\ \sqrt{8+x^2}\delta m^{\nu 2}_{rad}.
\end{eqnarray}

To get numerical estimations, let us fix $\vert x\vert$ = $\sqrt{2}$ corresponding to $\sin^22\theta_{12}$ = 0.8 and also fix $r^{\prime\prime}=1$ ($v = v^{\prime\prime}$) and $r^2$ = 1/9 ($v^\prime = v^{\prime\prime}/3$) corresponding to $\sin^2\theta_{23}$ = 0.95, where $v$ (=$v^{\prime\prime}\gg v^\prime$) $\approx$ $v_{weak}/\sqrt{2}$ to satisfy $v^2$ + $v^{\prime 2}$ + $v^{\prime\prime 2}$ = $v^2_{weak}$.  In the end, we derive $\sin^22\theta_{12}$ = 0.78 from reasonable assumptions on the couplings.  The conditions of Eq.(\ref{Eq:Couplings_Constraint}) in turn require
\begin{eqnarray}\label{Eq:CouplingEstimated1}
&&
\vert f_-/g_-\vert  \sim 3
\end{eqnarray}
to be consistent with $r^2$=1/9 and give the estimation of the Yukawa couplings to be:
\begin{eqnarray}\label{Eq:CouplingEstimated2}
&&
f_+ \sim g_+ \sim g_- \sim f_-/3 \sim 0.005.
\end{eqnarray}
The tree level mass of $m^\nu$ is estimated to be $\sim 0.03$ eV for $\Delta m^2_{atm} = 3 \times 10^{-3}$ eV$^2$ and $\delta m^\nu_{rad}$ = 3.2$\times 10^{-3}$ eV is obtained for $\Delta m^2_\odot = 4.5\times 10^{-5}$ eV$^2$.  Since $r^2$ in Eq.(\ref{Eq:Requirement2}) is almost saturated by $-s_\alpha$ for these values of $m^\nu$ and $\delta m^\nu_{rad}$, we observe that $s_\alpha \sim -1/9$.  The type II seesaw mechanism for $m^\nu$ yields an estimate of the mass of $s$: $m_s$ ($= \mu$) = 1.7$\times 10^{14} \times (\vert f^s_+\vert /e)$ GeV, where $e$ is the electromagnetic coupling.  From the expression of $\delta m^\nu_{e\mu}$ in Eq.(\ref{Eq:RadMassNuEntries}), we find that the estimation of $\delta m^\nu_{rad}$ yields

\begin{eqnarray}\label{Eq:CouplingValues}
&&
f^h_+\sim 2.3\times 10^{-7},
\end{eqnarray}
where $\mu_+$ = $m_\phi$ = $v_{weak}$ and $m^2_{h^+}$ = $10v^2_{weak}$ are used to compute the loop-factor of $P$.  From Eqs.(\ref{Eq:Requirement3}) and (\ref{Eq:CouplingRatio}), we also find that $x$ = $26\delta m^\nu_{\mu\tau}/27\delta m^\nu_{rad}$, which yields $\vert x \vert$ = $\sqrt{2}$ for $\vert\delta m^\nu_{\mu\tau}\vert/\delta m^\nu_{rad} = 1.47$, leading to 
\begin{eqnarray}\label{Eq:FixedParameters}
&&
f^h_-\mu_-/f^h_+\mu_+ = -9\sqrt{2}(1-r^2)x/52r=\pm 0.92,
\end{eqnarray}
from Eq.(\ref{Eq:CouplingRatio}) with $\vert \delta m^\nu_{e\mu}\vert \approx \delta m^\nu_{rad}$.  Finally, the masses of $m_{1,2,3}$ are predicted to be: 
\begin{eqnarray}\label{Eq:MassiveMass}
&&
\vert m_1 \vert = 2.8 \times 10^{-3}~{\rm eV},
\quad 
\vert m_2 \vert = 7.3 \times 10^{-3}~{\rm eV},
\quad
m_3 = 5.5 \times 10^{-2}~{\rm eV}.
\end{eqnarray}

It is remarkable to note that the result of these numerical estimates is consistent with the reasonable expectation of $\vert f^h_+\vert$=${\mathcal{O}}$($\vert f^h_-\vert$) and $\mu_+$=${\mathcal{O}}$($\mu_-$), but with neither $\vert f^h_+\vert\gg \vert f^h_-\vert$ nor $\vert f^h_+\vert\ll \vert f^h_-\vert$, to yield the large solar neutrino mixing.  This result should be contrasted with the requirement of ``inverse" hierarchy for the original Zee model \cite{InverseCoupling}.  If the relation of $f^h_+\sim f^h_-$ and $\mu_+\sim\mu_-$ is assumed, one finds that $\vert x\vert$ $\sim$ 1.53 for $r^2$=1/9 leading to $\sin^22\theta_{12}$ $\sim$ 0.78 in good agreement with the observed data. 

Summarizing our discussions, in the radiative mechanism based on the conservation of $L^\prime$ and the invariance under the $S_2$-transformation as well as under the discrete $Z_4$-transformation, we have demonstrated that the almost maximal atmospheric neutrino mixing is guaranteed by the $S_2$-symmetric coupling of $s$ to neutrinos and the large solar neutrino mixing is derived by the radiative effects only, where the tree-level contributions from $s$ vanish owing to the presence of $S_2$.  Our model spontaneously breaks $L$ and $L^\prime$ but preserves $L+L^\prime$, namely $L_e$.  This remaining $L_e$-conservation is used to select the Higgs interactions that include the key interactions for type II seesaw mechanism and radiative mechanism.  The massless Nambu-Goldstone boson associated with the spontaneous breakdown of $L-L^\prime$, namely $L_\mu+L_\tau$, can be removed by introducing a soft breaking such as $\phi^{\prime c \dagger}\phi^{\prime\prime}h^{+ \prime}$.  The model seems to suffer from the emergence of the dangerous flavor changing interactions that disturbs the well-established low-energy phenomenology of leptons because there are three Higgs scalars of $\phi$, $\phi^\prime$ and $\phi^{\prime\prime}$.  However, the explicit calculations show that the lepton sector has couplings to those Higgs scalars at most of order $m_\tau/v_{weak}$, which are shown to be sufficiently small to suppress these interactions to the phenomenologically consistent level.

In our scenario, properties of neutrino masses are summarized as follows:
\begin{enumerate}
	\item The smallness of neutrino masses is ensured by type II seesaw mechanism for atmospheric neutrinos and by radiative mechanism for solar neutrino neutrinos. 	\item The observed hierarchy of $\vert\Delta m^2_{atm}\vert\gg \vert\Delta m^2_\odot\vert$ is reproduced by the huge mass scale of $s$ of ${\mathcal{O}}$(10$^{14}$) GeV, which determines the democratic neutrino mass to be around 0.03 eV, and by the feeble couplings of $h^+$ ($h^{\prime +}$) to neutrinos of ${\mathcal{O}}$(10$^{-7}$), which determine the radiative neutrino mass to be around 0.003 eV.
\end{enumerate}
and neutrino mixings are explained as follows:
\begin{enumerate}
	\item The mixing angle of $\theta_{23}$ for atmospheric neutrinos is determined to be $t_{23}=(1-t_\alpha)/(1+t_\alpha)$ by $S_2$- and $Z_4$-symmetric tree-level mass terms without radiative effects ensuring $t_{23}\approx 1$ because $\vert \sin\alpha\vert\ll 1$ for the hierarchical $\mu$ and $\tau$ mass texture.  Radiative effects in turn give $t_{23}=-\delta m^\nu_{e\tau}/\delta m^\nu_{e\mu}=(1+r^2)/(1-r^2)$ with $r$=$\langle 0 \vert \phi^{\prime 0}\vert 0 \rangle/\langle 0 \vert \phi^{\prime\prime 0}\vert 0 \rangle$ fixing $f_-/g_- \sim r$ from Eq.(\ref{Eq:Couplings_Constraint}), which becomes consistent if $r^2 \sim -s_\alpha$ as in Eq.(\ref{Eq:Requirement2}).
	\item The mixing angle of $\theta_{12}$ for solar neutrinos is determined to be $\sin^22\theta_{12}=8/(8+x^2)$ by radiative mass terms, where $x$ = $\delta m^\nu_{\mu\tau}/\vert\delta m^\nu_{e\tau}\vert$, which is subject to the cancellation of the tree-level contributions in $x$ ensured by $S_2$. The ratio $\vert x \vert$ of ${\mathcal{O}}$(1) needed for the explanation of the large solar neutrino mixing can be realized by the requirement of $f^h_+\sim f^h_-$ and $\mu_+\sim\mu_-$, where $f^h_+$ ($f^h_-$) and $\mu_+$ ($\mu_-$) stand for $h^+$ ($h^{\prime +}$)-couplings, respectively, to leptons and to Higgs scalars.
\end{enumerate}
It should be noted that the hierarchical mass texture for $\mu$ and $\tau$ characterized by the finite mixing angle with $\sin^2\alpha$ ($\sim \sin^2\beta$) $\ll$ 1 is inevitable to be consistent with the almost atmospheric neutrino mixing.

We have also estimated various couplings by assuming the reasonable parameter setting based on $\langle 0 \vert \phi^0 \vert 0 \rangle\sim\langle 0 \vert \phi^{\prime\prime 0} \vert 0 \rangle$ and $\langle 0 \vert \phi^{\prime 0} \vert 0 \rangle$ $\sim$ $\langle 0 \vert \phi^{\prime\prime 0} \vert 0 \rangle/3$ or equivalently $s_\alpha\sim -1/9$, leading to $\sin^22\theta_{23}\sim 0.95$. The solar neutrino mixing parameter of $x$ is estimated to be $\vert x\vert$ = $26\vert\delta m^\nu_{\mu\tau}\vert/27\delta m^\nu_{rad}$ $\approx$ $\vert 13f^h_-\mu_-/6\sqrt{2}f^h_+\mu_+\vert$.  This estimation  yields $\vert\delta m^\nu_{\mu\tau}\vert/\delta m^\nu_{rad} = 1.47$ with $f^h_-\mu_-/f^h_+\mu_+ = \pm 0.92$ for $\vert x \vert$ = $\sqrt{2}$ corresponding to $\sin^22\theta_{12} = 0.8$ and implies the acceptable assumption of $f^h_+\sim f^h_-$ and $\mu_+\sim\mu_-$ giving the large solar neutrino mixing of $\sin^22\theta_{12}$ $\sim$ 0.78. In this respect, our mechanism provides a natural solution to the large solar neutrino mixing.  The presence of the permutation symmetry of $S_2$ for the $\mu$- and $\tau$-families in radiative neutrino mechanism enhances the significant deviation of $\sin^22\theta_{12}$ from unity as suggested by the latest data provided that the hierarchical mass texture is realized for $\mu$ and $\tau$.

\begin{center}
{\bf ACKNOWLEDGMENTS}
\end{center}

The authors are grateful to Y. Koide for useful discussions on the status of the Zee model.  One of the authors (M.Y.) also thanks the organizers and participants in the Summer Institute 2001 at FujiYoshida, Yamanashi, Japan, especially M. Bando and M. Tanimoto, for useful suggestions at the preliminary stage of this work.  The work of M.Y. is supported by the Grants-in-Aid for Scientific Research on Priority Areas A: ``Neutrino Oscillations and Their Origin" (No 12047223) from the Ministry of Education, Culture, Sports, Science, and Technology, Japan.

\vspace{4mm}
\centerline{TABLES}
\noindent
\begin{table}[ht]
    \caption{\label{Tab:QuantumNumber1}The lepton number ($L$), $L^\prime$ and $S_2$ and $Z_4$ for leptons and Higgs scalars, where $S_2=+ (-)$ denotes symmetric (antisymmetric) states.}
\begin{center}
\begin{tabular}{ccccccccccccc}\hline
&
$\psi_{eL}$
&
$e_R$
&
$\psi_{+L}$
&
$\psi_{-L}$
&
$\ell_{+R}$
&
$\ell_{-R}$
&
$\phi$
&
$\phi^\prime$
&
$\phi^{\prime\prime}$
&
$h^+$
&
$h^{\prime +}$
&
$s$
\\ \hline
$L$
&
1
&
1
&
1
&
1
&
1
&
1
&
0
&
0
&
0
&
$-2$
&
$-2$
&
$-2$
\\
$L^\prime$
&
1
&
1
&
$-1$
&
$-1$
&
$-1$
&
$-1$
&
0
&
0
&
0
&
0
&
2
&
2
\\
$S_2$
&
+
&
+
&
+
&
$-$
&
+
&
$-$
&
+
&
$-$
&
$-$
&
+
&
$-$
&
+
\\
$Z_4$
&
+
&
+
&
+
&
i
&
+
&
i
&
+
&
$-i$
&
$i$
&
+
&
$-i$
&
+
\\ \hline
\end{tabular}
\end{center}
\end{table}
\vspace*{-1cm}
\begin{table}[ht]
    \caption{\label{Tab:QuantumNumber2}$L$ and $L^\prime$ for Higgs interactions with $S_2$=$Z_4$=+.}
\begin{center}
\begin{tabular}{ccccccc}\hline
&
$\phi^{\prime c \dagger}\phi^{\prime\prime} h^{+\dagger}$ 
&
$\phi^{c \dagger}\phi^{\prime\prime} h^{\prime +\dagger}$ 
&
$\left( h^+h^+\right)^\dagger \det s$ 
&
$\phi^{\prime\prime \dagger} s\phi h^{\prime +\dagger}$, 
$\phi^\dagger s\phi^\prime h^{\prime +\dagger}$ 
&
$\phi^\dagger s\phi h^{+\dagger}$,
$\phi^{\prime\dagger} s\phi^\prime h^{+\dagger}$,
$\phi^{\prime\prime\dagger} s\phi^{\prime\prime} h^{+\dagger}$
&
$\phi^\dagger s\phi^c$,
$\phi^{\prime \dagger} s\phi^{\prime\prime c}$ 
\\ \hline
$L$
&
2 
&
2
&
0
&
0
&
0
&
$-2$
\\
$L^\prime$
&
0 
&
$-2$
&
4
&
0
&
2
&
2
\\ \hline
\end{tabular}
\end{center}
\end{table}

\noindent
\centerline{FIGURES}
\noindent
\noindent
\begin{figure}
\caption{One-loop diagrams for Majorana mass terms, where $i,j = e, \mu,\tau$ with $m,n=\mu,\tau$ and $M^\ell_0\left( \phi, \phi^\prime, \phi^{\prime\prime} \right)$ is defined by Eq.(\ref{Eq:OUR_M_ELL}).}
\label{Fig:OneLoop1}
\vspace{-7mm}
\end{figure}

\newpage
\centerline{\epsfbox{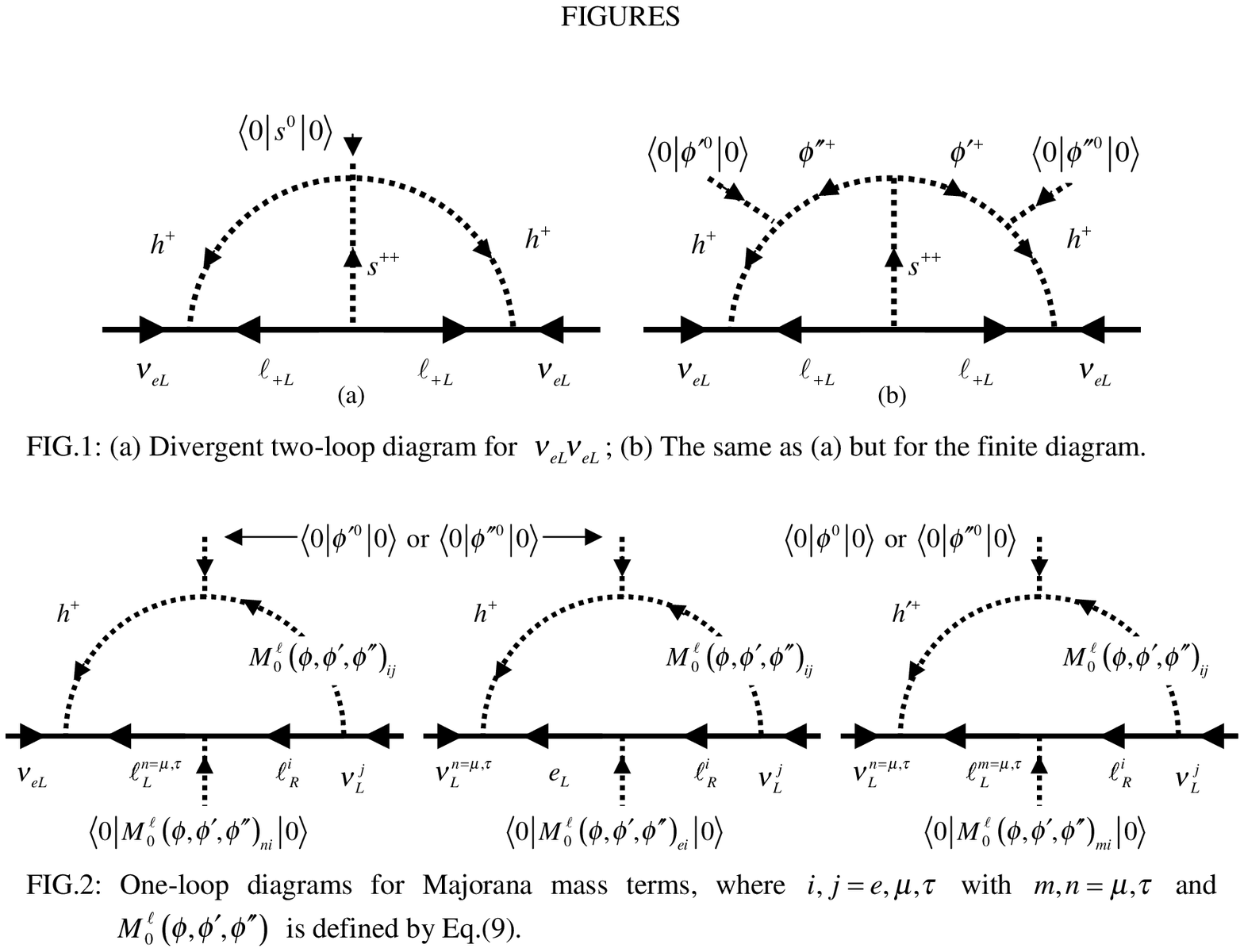}}


\begin{references}

%
\bibitem{EarlyMassive} 
    Z. Maki, M. Nakagawa and S. Sakata, \Journal{\PTP}{28}{870}{1962}. 
    See also  B. Pontecorvo, \Journal{\JETPUSSR}{34}{247}{1958};  
    B. Pontecorvo, \Journal{\ZETP}{53}{1717}{1967};
    V. Gribov and B. Pontecorvo, \Journal{\PLBOLD}{28B}{493}{1969}. 

%
\bibitem{Kamiokande} 
    Super-Kamiokande Collaboration, Y. Fukuda. et. al., \Journal{\PRL}{81}{1562}{1998};
    \Journal{\PLB}{433}{9}{1998}; \Journal{\PLB}{436}{33}{1998};
    N. Fornengo, M.C. Gonzalez-Garcia and J.W.F. Valle, \Journal{\NPB}{580}{58}{2000};
    T. Kajita and Y. Totsuka, \Journal{\RMP}{73}{85}{2001};
    K. Nishikawa, Talk given at 
    \textit{the Third International Workshop on Neutrino Factories based on Muon Storage Rings} (NuFACT'01), 
    24-30 May, Tsukuba, Japan (http://psux1.kek.jp/\textasciitilde nufact01/Docs/programs.html);
    C. Walter, Talk given at 
    \textit{the Third International Workshop on Neutrino Factories based on Muon Storage Rings} (NuFACT'01), 
    24-30 May, Tsukuba, Japan (http://psux1.kek.jp/\textasciitilde nufact01/Docs/agenda\_wg1.html).

%
\bibitem{SNO}
    SNO Collaboration, Q.R. Ahmad, et. al., \Journal{\PRL}{87}{07301}{2001}. 

%
\bibitem{K2K}
    K2K Collaboration, J.E. Hill, hep-ex/0110034. 

%
\bibitem{Seesaw} 
    T. Yanagida, in {\it Proceedings of the Workshop on Unified Theories and 
    Baryon Number in the Universe} edited by A. Sawada and A. Sugamoto 
    (KEK Report No.79-18, Tsukuba, 1979), p.95; \Journal{\PTP}{64}{1103}{1980};  
    M. Gell-Mann, P. Ramond and R. Slansky, in {\it Supergravity} edited by P. van 
    Nieuwenhuizen and D.Z. Freedmann (North-Holland, Amsterdam 1979), p.315; 
    R.N. Mohapatra and G. Senjanovi\'{c}, \Journal{\PRL}{44}{912}{1980}.

%
\bibitem{Zee}
    A. Zee, \Journal{\PLBOLD}{93B}{389}{1980}; \Journal{\PLBOLD}{161B}{141}{1985};
    L. Wolfenstein, \Journal{\NPB}{175}{93}{1980};
    S.T. Petcov, \Journal{\PLBOLD}{115B}{401}{1982}.

%
\bibitem{Babu}
    A. Zee, Nucl. Phys. {\bf 264B} (1986) 99; 
    K. S. Babu, Phys.  Lett. B {\bf 203} (1988) 132; 
    D. Chang, W-Y.Keung and P.B. Pal, Phys. Rev. Lett. {\bf 61} (1988) 2420; 
    J. Schechter and J.W.F. Valle, Phys. Lett. B {\bf 286} (1992) 321.

%
\bibitem{RecentAtmAnalysis}
    See for example, 
    J.W.F. Valle, hep-ph/0104085;
    E. Lisi, Talk given at 
    \textit{the Third International Workshop on Neutrino Factories based on Muon Storage Rings} (NuFACT'01), 
    24-30 May, Tsukuba, Japan (http://psux1.kek.jp/\textasciitilde nufact01/ Docs/agenda\_wg1.html).

%
\bibitem{RecentSK}
    D. Wark, Talk given at 
    \textit{the Third International Workshop on Neutrino Factories based on Muon Storage Rings} (NuFACT'01), 
    24-30 May, Tsukuba, Japan (http://psux1.kek.jp/\textasciitilde nufact01/Docs/programs.html);
    Y. Fukuda, Talk given at 
    \textit{the Third International Workshop on Neutrino Factories based on Muon Storage Rings} (NuFACT'01), 
    24-30 May, Tsukuba, Japan (http://psux1.kek.jp/\textasciitilde nufact01/Docs/agenda\_wg1.html).

%
\bibitem{BimaximalMixing}
    D. V. Ahluwalia, \Journal{\MPL}{13}{2249}{1998};
    V. Barger, P. Pakvasa, T.J. Weiler and K. Whisnant, \Journal{\PLB}{437}{107}{1998};
    A. Baltz, A.S. Goldhaber and M. Goldhaber, \Journal{\PRL}{81}{5730}{1998};
    M. Jezabek and Y. Sumino, \Journal{\PLB}{440}{327}{1998};
    R.N. Mohapatra and S. Nussinov, \Journal{\PLB}{441}{299}{1998};
    Y. Nomura and T. Yanagida, \Journal{\PRD}{59}{017303}{1999};
    I. Starcu and D.V.Ahluwalia, \Journal{\PLB}{460}{431}{1999};
    Q. Shafi and Z. Tavartkiladze, \Journal{\PLB}{451}{129}{1999};
    \Journal{\PLB}{482}{145}{2000};
    C.H. Albright and S.M. Barr, \Journal{\PLB}{461}{218}{1999};
    H. Georgi and S.L. Glashow, \Journal{\PRD}{61}{097301}{2000};
    R.N. Mohapatra, A. P\'{e}rez-Lorenzana and C. A. de S. Pires, \Journal{\PLB}{474}{355}{2000}.

%
\bibitem{NearlyBiMaximal} 
    H. Fritzsch and Z.Z. Xing, \Journal{\PLB}{372}{265}{1996}; \Journal{\PLB}{440}{313}{1998};
    M. Fukugita, M. Tanimoto and T. Yanagida, \Journal{\PRD}{57}{4429}{1998}; 
    M. Tanimoto, \Journal{\PRD}{59}{017304}{1999}.

%
\bibitem{Useful} 
    C. Jarlskog, M. Matsuda, S. Skadhauge and M. Tanimoto, \Journal{\PLB}{449}{240}{1999};
    P.H. Frampton and S.L. Glashow, \Journal{\PLB}{461}{95}{1999};
    A.S. Joshipura and S.D. Rindani, Phys. Lett. B {\bf 464} (1999) 239; 
    D. Chang and A. Zee, \Journal{\PRD}{61}{071303}{2000}.

\bibitem{EarlierLprime}
    S.T. Petcov, \Journal{\PLBOLD}{110B}{245}{1982};
    C.N. Leung and S.T. Petcov, \Journal{\PLBOLD}{125B}{461}{1983};
    A. Zee, \Journal{\PLBOLD}{161B}{141}{1985}.
%
\bibitem{Lprime}
    R. Barbieri, L.J. Hall, D. Smith, N.J. Weiner and A. Strumia, \Journal{\JHEP}{12}{017}{1998}.

%
\bibitem{RecentSolarAnalysis}
    See for example, 
    C. Pe\~{n}a-Garay, Talk given at 
    \textit{the Third International Workshop on Neutrino Factories based on Muon Storage Rings} (NuFACT'01), 
    24-30 May, Tsukuba, Japan (http://psux1.kek.jp/\textasciitilde nufact01/Docs/agenda\_wg1.html);
    J.N. Bahcall, M.C. Gonzalez-Garcia and C. Pe\~{n}a-Garay, \Journal{JHEP}{08}{014}{2001};
	V. Barger, D. Marfatia and K. Whisnant, hep-ph/0106207;
	P.I. Krastev and A.Yu. Smirnov, hep-ph/0108177.

%
\bibitem{ZeeMaximal}
    For the explicit demonstration, see P.H. Frampton and S.L. Glashow, in Ref.\cite{Useful};
    Y. Koide, \Journal{\PRD}{64}{077301}{2001}.

%
\bibitem{Previous}
    T. Kitabayashi and M. Yasu\`{e}, Talk given at \textit{Summer Institute 2001} (SI2001), 13-20 August, FujiYoshida, Yamanashi, Japan, to appear in the Proceedings.

%
\bibitem{ZeeLMA}
    K.R.S. Balaji, W. Grimus and T. Schwetz, \Journal{\PLB}{508}{301}{2001}.


%
\bibitem{DiscreteSym}
    W. Grimus and L. Lavoura, hep-ph/0105212 and 0110041.

%
\bibitem{Triplet}
    E. Ma and U. Sarkar, \Journal{\PRL}{80}{5716}{1998}.

%
\bibitem{Type2SeeSaw}
   R.N. Mohapatra and G. Senjanovi\'{c}, \Journal{\PRD}{23}{165}{1981};
   C. Wetterich, \Journal{\NPB}{187}{343}{1981}.
   For a model with the similar scenario to ours in the type II seesaw 
   mechanism, see R.N. Mohapatra, A. P\'{e}rez-Lorenzana and 
   C.A.de S. Pires, in Ref.\cite{BimaximalMixing}.

%
\bibitem{MassMatrix}
   E. Ma. M. Raidal and U. Sarkar, \Journal{\PRL}{85}{3769}{2000};
   S.K. Kang and C.S. Kim, \Journal{\PRD}{63}{113010}{2001};
   C.S. Lam, \Journal{\PLB}{507}{214}{2001}.
   See also, I. Dorsner and S.M. Barr, hep-ph/0108168.

%
\bibitem{Moha} 
   See for example, R.N. Mohapatra, \Journal{\NPSUPPL}{91}{313}{2001}.

%
\bibitem{Permutation}
    For leptons with $S_3$, see for example, M. Fukugita, M. Tanimoto and T. Yanagida, in Ref.\cite{NearlyBiMaximal};
    M. Tanimoto, in Ref.\cite{NearlyBiMaximal};
    R.N. Mohapatra, A. P\'{e}rez-Lorenzana and C. A. de S. Pires, in Ref.\cite{BimaximalMixing}. For quarks with $S_3$, see H. Harari, H. Haut and J. Weyers, \Journal{\PLBOLD}{78B}{459}{1978}.

%
\bibitem{Democratic}
See for example, H. Harari, H. Haut and J. Weyers, in Ref.\cite{Permutation};
    Y. Koide, \Journal{\PRD}{28}{252}{1983}; \Journal{\PRD}{39}{1391}{1989};
    P. Kaus and S. Meskov, \Journal{\MPL}{3}{1251}{1988};
    M. Tanimoto, \Journal{\PRD}{41}{1589}{1990};
    G.C. Branco, J.I. Silva-Marcos and M.N. Rebelo, \Journal{\PLB}{237}{446}{1990};
    H. Fritzsch and J. Plankl, \Journal{\PLB}{237}{451}{1990}.

%
\bibitem{Hierarchical} 
    H. Fritzsch, \Journal{\PLBOLD}{70B}{436}{1977}; \Journal{\PLBOLD}{73B}{317}{1978};
    S. Weinberg, \Journal{\TNYAS}{38}{185}{1977};
    F. Wilczek and A. Zee, \Journal{\PLBOLD}{70B}{418}{1977}.

%
\bibitem{Data}
    Particle Data Group, D.E. Groom et al., \Journal{\EPJ}{15}{1}{2000}.

%
\bibitem{h_phenom2}
   See for example, E. Mitsuda and K. Sasaki, \Journal{\PLB}{516}{47}{2001};
   A. Ghosal, Y. Koide and H. Fusaoka, \Journal{\PRD}{64}{053012}{2001}.

%
\bibitem{InverseCoupling} 
   C. Jarlskog, M. Matsuda, S. Skadhauge and M. Tanimoto, in Ref.\cite{Useful}.
   See also, Y. Koide and A. Ghosal, \Journal{\PRD}{63}{037301}{2001}.

\end{references}
\end{document}